# HCI and Educational Metrics as Tools for VLE Evaluation

Vita Hinze-Hoare

University of Southampton

School of Electronics and Computer Science

# Table of Contents

---

# HCI and Educational Metrics as Tools for VLE Evaluation

*"It's not what you teach but what they learn that matters"*

Ancient Arabian Proverb

Abstract

*Under consideration are the general set of Human computer Interaction (HCI) and Educational principles from prominent authors in the field and the construction of a system for evaluating Virtual Learning Environments (VLEs) with respect to the application of these HCI and Educational Principles. A frequency analysis of principles is used to obtain the most significant set. Metrics are devised to provide objective measures of these principles and a consistent testing regime is introduced. These principles are used to analyse the University VLE Blackboard. An open source VLE is also constructed with similar content to Blackboard courses so that a systematic comparison can be made. HCI and Educational metrics are determined for each VLE.*

## 1. Introduction

A well-known problem[1] with the evaluation of educational virtual learning environments (VLEs) is the lack of a clear objective assessment framework having a wide recognition. This means that there is an issue over the best way of evaluating their effectiveness on both, sound educational principles and on Human Computer Interaction or design principles.

There is at present no common agreement as to the order of ranking for either HCI or Educational standards, which have always lacked objective measurement[2] and universal acceptance. Fragmented and diverse sets of principles originate from many sources each with their own proponents and champions.

It is the aim of this project to move towards some objective standard by which to gauge VLEs, and ultimately to provide a single overall index measure (essentially a score out of 10) for both usability and educational worth. To this end an HCI index will be constructed for general usability comparison, and a separate educational index (EDI) will be designed to provide a measure of educational quality. These single metrics will allow a direct measure of VLEs.

The resultant indices will be calculated for two VLEs: First the Blackboard VLE which is used to teach ITO course units and secondly an open source VLE. As far as possible, the open source VLE will carry the same content as the Blackboard VLE to allow a comparison of the VLE structure and operation rather than its content.

Usability statistics are to be obtained from a set of standard users.

### 1.1. Background

The Dearing Report (1997)[3] focussed on developments in Internet and Intranet delivery of learning objectives and the support of the lecturing environment, and required Higher Education Establishments to achieve a vision of "a learning society".

This has been adopted nationally by all Institutions of Higher Education and increasing numbers of teaching staff are making their lecture notes and reading lists available through this medium.

The use of Virtual Learning Environments (VLEs) and to a more limited extend Managed Learning Environments (MLEs) are becoming more than common practice at all levels of education. It is now seen as essential for each institution to consider moving a proportion of their teaching onto a VLE/MLE delivery structure, and that the Government through organisations such as FEDA, JISC etc. are giving the lead in steering education in this direction. Digital delivery not only provides alternative teaching and learning strategies but also interleaves well with the widening participation agenda also promoted by the Government. Furthermore as financially hard-pressed educational establishments are looking for new markets to deliver their courses, the VLE provides an almost ideal or even magical solution to the delivery of remote learning.

## 1.2. Aims and Purpose

The proliferation of VLEs has led to a number of surveys and comparative reviews which seek to provide an objective method of determining the relative usefulness and suitability of the software. See for example Bruce Landon[4], University of British Columbia, for what is claimed to be the most comprehensive comparative evaluation. This is an online extensible review to which users can add their own comments.

However, as useful as these comparative evaluations are, it is clear that they do not distinguish a great deal between different systems according to Sandy Britain 1998[5]. This is because the question of suitability is not merely about which system offers the most features or the best functionality. This kind of analysis says nothing about how easy or difficult it is to carry out a task nor whether the VLE structure promotes or works against current theories of learning and teaching.

It is suggested that rather than a feature comparison, what is most needed is a holistic approach to the evaluation of these packages, which particularly takes as a set of standards the ascertaining of the educational principles around which the system functions, and how well it fulfils the needs of the learner.

Two key considerations have been identified:

- VLEs should provide opportunities to improve teaching and learning based on pedagogical perspectives and -

- VLEs should reduce the amount of work required to learn and be founded on sound ergonomic and usability  (HCI) principles.

Evaluating VLEs needs to be done at different levels. On the higher strategic level of educational theory and its application to the design and delivery of teaching and learning content needs evaluation. In addition, on the lower operational level an analysis of the usability characteristics based upon standardised HCI design principles also needs to be carried out.

| Level 1 | Strategic | Application of Educational Principles |
| Level 2 | Operational | Application of HCI Principles |

## 1.3. Developing an evaluation strategy

The four principle methods of evaluating involve:

### 1.3.1. The quick and dirty evaluation.
These can be done at any stage and the emphasis on fast input rather than carefully documented findings. It is called quick and dirty because it is done in a short space of time and the feedback may not be 100% reliable with the emphasis on the descriptive and the informal such as may be obtained from an action research approach. Dick, B. (2002)[6]

### 1.3.2. Usability testing
This is the approach dominant in the 1980's, and involves measuring the users' performance on carefully prepared tasks. Users may be recorded on video or logging their interactions via software. This is a strongly controlled environment often taking place in laboratory like conditions. "Whiteside et al 1988 [7]

### 1.3.3. Field studies.
These are done in natural settings with the aim of increasing the understanding of how users behave in a natural environment and how the technology really impacts upon true-life conditions. Field studies are often used for the evaluation of technology. Bly 1997[8] Some times the data will take the form of events or conversations, interviews or even direct usage analysis. Content analysis where the data is analysed into content categories is a common feature.

### 1.3.4. Predictive Evaluation
Here experts apply their knowledge of typical users to try to predict problems that occur with usability. This kind of evaluation does not require users at all. This makes the process quick, inexpensive and attractive to companies. However limitations of the results can be significant and results can often be inappropriate and missing vital factors. Great care is needed in setting up this kind of evaluation and a great deal hinges upon the expertise of the analyst.

## 1.4. Choosing an evaluation method
The field study approach will be adopted in this project. The reasons for this are based on a number of factors including the inadequacy of some approaches for the specific task at hand and the inability or lack of resources. The quick and dirty method has been rejected as being too unreliable for this project, which requires a reasonable degree of accuracy and confidence in the results. The usability testing method with its emphasis on laboratory techniques was not required as facilities were not readily available to all participants and environmental control was not needed. The predictive approach has also been discarded on the basis that expert reviews would require the use of consultants and was inappropriate in this instance. However, field study evaluation, which looks at the natural behaviours of the student, fits well with the purposes and goals of this enquiry.

## 2. Evaluation of VLEs using the HCI and Educational Frameworks

### 2.1. Examining possible Frameworks and Approaches

One approach that might be taken in the HCI area would be the adoption of a particular author's principles and construct a framework from that. However this leads to the objection that one particular author cannot be promoted over another. This approach would lead to a very subjective framework, which would not be widely accepted.

An improvement on this approach would be the construction of a heterogeneous framework built upon the principles of many authors. However, this leads to the similar objection of promoting one group of authors over another and leads to the further complication of how to organise the principles into a consistent hierarchy.

A third approach which seeks to find both an objective and widely accepted set of principles might be best obtained from an analysis of the citation frequency of the authors in the field, and a ranking of those authors in order to obtain that group which is most widely accepted. A further analysis of the literature produced by this specific group of authors would lead to the listing of the HCI principles most commonly promoted by performing a frequency analysis of principles. This would enable to establish a definitive list of the most widely accepted principles into which we can construct the framework.

In the area of educational principles the approach is not quite so straightforward as competing pedagogical frameworks are mutually exclusive which means that only one theory can be adopted at the expense of the others. This is critically different from the HCI position were different principles proposed by different authors are mutually inclusive and a principle from one author can be combined into the same framework as a principle from another author. The consequence of this is that one educational theory will need to be adopted over and above other educational theories.

Sandy Britain 1998 working on behalf of JISC has evaluated VLEs on the basis of the conversational framework developed by Laurillard 1993[9]. This provides a medium for supporting constructivist and conversational approaches to learning with the central emphasis being placed upon dialogue between student and teacher. The weakness of this work is in the limited acceptance of Laurillard's methodology.

Alternative approaches have been done using other models including a cybernetic model which centres upon the structure of the learning environment and how VLEs allow for the management of groups of learners and the interactions between people in different roles and the mechanisms which are used to integrate and organise content, timetables etc. The weakness of this particular area is that it deals more with the mechanism rather than the pedagogy.

It is necessary to adopt the most widely accepted educational theory available today. This will ensure that we have the widest recognition possible for the educational metric.

## 2.2. Formation of the appropriate questions

Questionnaires are a well-established technique for collecting users opinions. It is important to ensure that the questions are clearly worded and data can be analysed efficiently. The advantage of questionnaires is that they can be distributed to a large number of people and hence provide evidence of a wide population addressed. They can also be completed in a shorter time than an interview. In addition questionnaires can be designed to be either open ended or closed. The first type arouses the user to comment on anything and everything that he wishes, whereas the second type is closely contained and ranges only over those areas the questioner is interested in. The answers can also be predefined in the form of a multiple-choice format.

## 2.3. Questionnaire Design

In designing the questionnaire it was important to relate the questions to the principles being examined. This can be done in a multiplicity of forms enabling cross checking of answers against what amounts to the same question in a different format. By opting for a multiple choice questionnaire it is possible to provide a closer statistical analysis of student requirements and therefore this method was adopted. In addition it was further decided that in the case of VLE comparison it would be appropriate to ensure that all the answers provided consistent comparability to avoid any bias present within the answer framework. For this reason every question had a choice of five possible answers which allowed the user to rate the features on a scale that was mostly one VLE leading to the other VLE at the opposite end of the scale. This is the Likert scale (Oppenheim 1992) (pp401 Preece Interaction Design) used for measuring opinions, attitudes and beliefs and consequently user satisfaction.

### 2.3.1. Absolute Questions and Relative Questions

The two approaches which can be taken are firstly the determination of a single metric for an individual Website, or secondly the construction of a metric for the purposes of comparing two different websites. In the first instance the set of questions, which asks the user to rate in absolute terms using values from 1 to 5 (following Likert see Oppenheim 1992)[10] would be appropriate. In the second case the set of questions would be designed to compare the two VLEs using values, which related one Website in comparison to another. The following five categories have been used:

- Mostly VLE 1
- VLE 1 more than VLE 2
- VLE1 and VLE2 equally
- VLE2 more than VLE1

- Mostly VLE 2

These five categories correspond exactly to the Likert values of 1 to 5. For this reason it was important to keep the answers the same for each multiple choice question. This not only provided consistency of approach but also provided as far as possible any elimination of bias in the phrasing of the questions or the choosing of the answers.

The advantage of the first approach using absolute questions is that it provides individual metrics for each site. The advantage of the second is speed and efficiency, as it only requires the completion of a single questionnaire to compare the two sites.

In this project a single questionnaire is completed by the user to compare two VLEs. The relative question approach has been used to reduce questionnaire load on the user panel and provide a quicker and more efficient survey.

Although this provides essentially only a comparison index, it is still possible to give an indication of absolute values by the separation of the scores obtained. However it should be understood that HCI and educational indices obtained in this way can only be seen as providing an indicative score for illustration only. For full absolute indices individual questionnaires for each site must be done separately.

## 2.4. Survey Participants
Participants were required to have access to both VLEs. Whereas access would be provided to the Open Source VLE via a Website set up specifically for this purpose, it was important that Users had access to Blackboard in their own right. This limited the survey participants to the students of Southampton University for all practical purposes.

A number of methods were used to contact and recruit this category of participants.

- In the first instance the students of INFO 3001 were contacted via Dr.Mike Wald.
- Secondly the third year students of ITO were contacted via email directly.
- Thirdly Web page holders on the Blackboard system were sent direct emails.
- Fourthly a reminder was sent via email to the 3rd year ITO students.
- Fifthly students on the Multimedia COMP 3013 course were also contacted via email.
- Lastly students of COMP 3024 were contacted via Dr. David Argles

Participants were selected from Student Blackboard users rather than Lecturers because the aim was to test the educational principles involved in the learning process and not the administrative processes of managing the course on the VLE.

### 2.4.1. Representative Sample

Dumas and Reditch 1999 have established that five to twelve testers are a sufficient sample for on-line usability testing. However, the more testers there are the more representative the findings will be across the user population. (Preece Rogers & Sharpe 2002) pp441

### 2.4.2. Test Procedure

The procedure involved a field test of participants exploring the two systems in their own time and under their own direction. They were free to pursue and encouraged to examine all areas of the websites. No prescribed tests were used to enable users to concentrate on those areas which they were most interested in and which they would find most useful.

The danger of this field test approach is that performance differences occurring amongst participants might be attributed to different procedures. However, it was felt that this field test approach would best represent the natural responses of the users. Each participant was asked to work through 34 on-line questions, which, on average, took approximately six minutes to complete. No time limit was set for the completion of the questionnaire.

### 2.4.3. Data Collection

The questions were designed and then built into the Moodle VLE, which enabled rapid and automatic data collection. Each user's answers were automatically logged and sorted by the system. The data collected contained the following information:

- Start date and time
- Time for completion
- Users name
- Individual student answers for each question

### 2.4.4. Data Analysis

Some of the data analysis is performed automatically by the features of the Moodle VLE survey. This included:

- Accumulative percentage of answers for each question
- Accumulative rating of Blackboard versus Moodle on a percentage scale for each user
- Accumulated rating of Blackboard versus Moodle on a percentage scale for each question

Additional analysis was performed following the exporting the data into excel. This enabled comparison charts and graphs to be constructed as well as additional statistical analyses. A detailed focus on such things as splitting of the results into the HCI and Educational Question Components was performed, and in turn a further splitting of the results to produce an overall view of each of the Educational and HCI principles involved.

## 3. HCI Principles

The first task was to determine a set of principles that could be confidently represented as the most significant and widely supported in the field.

### 3.1. Methodology

It was deemed that the most cited authors would constitute the most highly respected in the field and therefore we began an analysis of the citation frequency HCI authors. In this we were helped by the HCI bibliography Website which stores a citation frequency database.[11] In order obtain the most significant HCI principles the following research program was devised.

#### 3.1.1. Researching the authors

A list of the most frequent authors (10 or more publications) in the HCI Bibliography, starting in December 1998, was generated from the author fields in the database. Although the HCI bibliography is not exhaustive it is nevertheless indicative of author citation frequency and as such establishes a base line for the specification of key research areas. The results from this website research are as follows:

| POSITION | AUTHOR | CITATION FRQ. |
|----------|--------|---------------|
| 1 | Ben Shneiderman | 165 |
| 2 | John M. Carroll | 128 |
| 3 | Steven Pemberton | 117 |
| 4 | Jakob Nielsen | 117 |
| 5 | Brad Myers | 101 |
| 6 | Gavriel,G Salvendy | 98 |
| 7 | Gary Perlman | 94 |
| 8 | Tom Stewart | 90 |

In addition to these most frequently cited authors there is an additional list of authors that have been used either because they have produced a comprehensive summary review of HCI principles or have written recent and widely used textbooks on the subject. These include Ben Schneiderman, Jakob Nielsen, Donald Norman, Brad Myers, Jenny Preece, Jef Raskin, Dix, Finlay, Abowd, Beale and others. By combining both sets of authors a final comprehensive list has been produced for the bibliography.

### 3.2. Researching the literature

An analysis table was drawn up; see Appendix 1, and the principles identified. Commonality was identified and broad categories outlined. A frequency analysis of the principles was undertaken and all principles with scores of 6 and above were accepted as the most significant. There were 8 of these listed below.

### 3.3. The Eight HCI Principles

| 1 | Familiarity | 12 |
|---|---|---|
| 2 | Consistency | 12 |
| 3 | Forward Error Recovery | 9 |
| 4 | Substitutivity | 8 |
| 5 | Dialogue Initiative | 7 |
| 6 | Task Migratablility | 7 |
| 7 | Responsiveness | 6 |
| 8 | Customisability | 6 |

### 3.4. Constructing the HCI Evaluation Framework

Each of the eight principles finds representation in a set of questions constructed specifically for the purpose of drawing out the users views on the principle concerned.

#### 3.4.1. Familiarity

This is the degree to which the user's own real world personal experience and knowledge can be drawn upon to provide an insight into the workings of the new system. The familiarity of a user is a measure of the correlation between their existing knowledge and the knowledge required to operate the new system. To a large extend familiarity has its first impact with the users' initial impression of the system and the way it is first perceived and whether the user can therefore determine operational methods from his own prior experience. If this is possible this greatly cuts down the learning time and the amount of new knowledge that needs to be gained. The term familiarity is proposed by Dix et al (1992) but is referred to by other authors under different terms for example Jordan et al (1991)[12] refers to familiarity as guessability. Schneiderman (1998) and Preece (1994) each refer to familiarity in terms of the reduction of cognitive load. Other terminological synonyms can be found in Appendix1.

This was the most quoted principle amongst all HCI authors and as such gained a rating of 12. In order to measure the degree by which the user was familiar with the operation of the VLE four questions were used

*Q6 Which site gave you the responses you expected?*
*Q29 Which site worked according to your expectation?*
*Q33 At which site did you feel most at home?*
*Q9 Which VLE do you think most students would use if given the choice*

#### 3.4.2. Consistency

Consistency, according to Dix et al (1992) relates to the likeness in behaviour arising from similar situations or similar task objectives. He also thinks that this is probably the most widely mentioned principle in the literature on user interface design. With this we can provide the statistical support from **Appendix 1** where this principle comes out as joint first place with familiarity. It is considered of vital importance that the user has a consistent interface.

However, there is an intrinsic difficulty in defining the nature of consistency, which can take many forms. Consistency is relative to a particular area for example one can speak of consistency of mouse movements, menu structures, response etc. Whereas familiarity can be considered as "consistency with respect to personal experience" this consistency is one with respect to "internal similarity of appearance and behaviour".

This principle shared the top slot with familiarity, also with a weighting of 12. This was tested by four questions.

**Q13** *Which site seemed more consistent in use?*
**Q31** *Which site was easiest to find your way around?*
**Q17** *Which site would you least want to change to make it more usable?*
**Q19** *Which was the easiest one to understand*

### 3.4.3. Forward Error Recovery
This is the ability of users recovering from their errors, which they invariably make. There are two directions in which recovery can occur both forward and backward. Forward error recovery involves the prevention of errors. Backward error recovery concerns the easy reversal of erroneous actions. The latter is usually concerned with the users actions and is initiated by the user. The former is one, which should be engineered into the system and initiated by the system. In this sense recoverability is connected to fault tolerance, reliability and dependability. Ken Maxwell (2001) considers this basic usability a level one priority, which he calls error protection. Jeff Raskin (2000) rates this as part of his first law of interface design, which states, "a computer shall not harm your work or through inaction allow your work to come to harm".

**Q8** *Which site allowed you to recover more easily from mistakes?*
**Q34** *Which site gave you the least operating problems?*

### 3.4.4. Substitutivity
This concerns the ability of the user to enter the same value, or perform the same action in different ways according to his own personal preference. For example a user might wish to enter values in either inches or centimetres, or he may wish to open a program with the mouse or with the keyboard. This input Substitutivity contributes towards an overall flexible HCI structure, which allows the user to chose whichever he considers most suitable. Schneiderman (1998) and Preece (1994) provide a specific example of providing shortcuts as an alternative.

This is the ability of the interface to provide multiple methods for performing the same task.

**Q23** *Which site gave you the better choice in the way to do what you wanted?*
**Q20** *Which site gave you the most options for task completion*

### 3.4.5. Dialogue Initiative
This allows the user the freedom from artificial constraints on the input dialogue boxes. When humans communicate with computers and the dialogue is set up it is important to ensure that the human partner has the initiative in the conversation. If the system initiate all dialogues and the user simply responds, this is called "system preemptive". On the other hand if the user is free to initiate actions then this is called "user preemptive". An effective system is one, which maximizes the users ability to preempt the system, and minimizes the systems ability to preempt the user. This is also referred to as allowing input flexibility, Preece (1994).

The dialogue between the user and the interface needs to be unambiguous speedy and effective. This is often conducted through dialogue boxes.

*Q4* *Which site had the most informative dialogue boxes?*

### 3.4.6. Task Migratablility
This concerns the transfer of control for executing tasks between the system and the user. Checking the spelling of a document is a good example. The user can quite easily check the spelling for himself by the use of a dictionary. However the task is made considerably easier if it can be passed over to the system to perform with simple checks made by the user as to the acceptable spelling i.e. the difference between US and British Dictionaries. This is an ideal task for automation. However, it is not desirable to leave it entirely in the hands of the computer as dictionaries are limited and therefore the task needs to be handed over to the user at those complex points where the system cannot cope. Ken Maxwell (2001) talks of this as level two collaborative organisational interaction which he considers being a high level of HCI interaction.

This is the ability of the interface to hand the task over to the user so that the initiative rests with the human side of the interaction. This can be measured by the degree of performance available through the use of unfamiliar tasks.

*Q11* *Which site was the easiest to use for unfamiliar tasks*

### 3.4.7. Responsiveness
This measures the rate of communication between the user and the system. It covers such areas as simple response time to keys pressed or to actions executed but also concerns the quality of the feedback provided by the system and the appropriateness of the responses that are made. For instance a computer system that takes 10 seconds to respond to a simple click on a toolbar button will rapidly lead to user frustration. On the other hand a system that provides rapid and appropriate responses, including both audio and visual cues will be well received by the user. Jeff Raskin (2000) writes of this as his second law of the computer interface which he defines as "a computer shall not waste your time...".

This is the immediacy and feel that the interface provides to the user and is an important subjective criterion of perceived efficiency.

*Q26* *Which site has the best look and feel?*

### 3.4.8. Customisability
This is ability of the user to modify the interface. This is sometimes known as adaptability and allows different users to adapt the interface according to their own level and style of interaction. This customisation can be enhanced by providing the user with various levels of interaction. At the lowest end it may merely involve the altering of colours and styles while at the upper end an application may allow the user to develop their own macros or even write their own VB modules. This allows for the elimination of repetitive tasks or for the construction of a personally more suitable environment to the user.

This involved the ability of the user to alter their environment to a certain extent and had a weighting of 5.

**Q18** *Which site is the most easily customisable?*
**Q1** *Which site do you prefer to work with best?*

This provided a total of 17 questions for the HCI evaluation framework.

## 3.5. The HCI Metric
The method used to create the HCI metric involves the scoring of the eight HCI principles chosen by users during the online field survey and the conversion of those scores into a single figure index ranging from 1 to 10.

Each of the eight HCI principles was evaluated using between one and four distinct questions for each principle making a total of 17 questions. Each question had a score rated from 1 to 5.

Stage one involved the taking of the average score for the total of the seventeen questions. Stage two involves multiplying that score by a factor of 2 to obtain a result out of ten.

HCI index = Average (Likert score for each HCI Question) x 2

In the case of relative comparison of two VLEs the Likert score is again applied uniformly to each of the answers. This is possible because all the comparative answers are identical and the same formulae will follow. However, the interpretation of the score would be that a score of 1 would indicate a 100% preference for VLE 1 while a score of 10 would indicate 100% preference for VLE 2. In a similar way a score of 3 would indicate a 50% preference for each VLE. And similarly intermediate values would indicate the proportional preference for the corresponding VLE.

The evaluation that is to be conducted for the purposes of this report involves this comparison set of questions, which are calculated automatically by the on-line survey mechanisms built into the Moodle Website. This was constructed in percentage terms in such a way that a score of 1 was rated as a value of 0% in favour of VLE 1, a score of 2 was valued at 25% in favour of VLE1 etc.

| Likert Score | VLE 1 | VLE 2 |
|---|---|---|
| 1 | 0% | 100% |
| 2 | 25% | 75% |
| 3 | 50% | 50% |
| 4 | 75% | 25% |
| 5 | 100% | 0% |

The VLE automatically constructed a value based on the table values given above allowing a relative HCI Index to be provided automatically.

## 4. Educational Principles

E-learning courses should incorporate methods that are based on sound (educational) research. Clark & Mayer (2003) pp 42. Much of the principle work in educational theory is found in books rather than original papers and has lead to a different approach from that adopted for the HCI authors.

At the outset an analysis was performed on educational principles following a similar method of that adopted for the HCI principles. See Appendix 1 It was discovered that whereas the principles of the HCI were **mutually inclusive** and could therefore be adopted in conjunction with each other, the Educational principles from different educational theorists were **mutually exclusive** and could not be mixed and matched in the same way. Each Educational principle was constructed as part of the edifice of a single educational theory and could not be removed and placed with a different principle in a different theory. A new approach was necessary and it was decided to adopt instead the most prevailing educational theory today from a single theorist.

We have adopted an approach that has centred upon a particular framework from one of the most highly cited educational theorists today, namely Jerome Bruner[13] together with additional support based upon the work of Vygotsky. Vygotsky was influenced by Piaget who was influential in the Graphical User Interface creation.

### 4.1. Bruner's Constructivism

For Bruner, "*learning is an active process in which learners construct new ideas or concepts*"[10]

According to Martin Dougiamas the creator of Moodle, "*Constructivism is building on knowledge known by the student. Education is Student centred, Students have to construct knowledge themselves*" (Dougiamas, M.1998).

Bruner's educational theory maintains that the prior knowledge of the learner is the essential element in constructing new knowledge. This is based upon Piaget's ideas that knowledge is actively constructed and not passively received.

In one of the seminal works on educational theory, Jerome Bruner's "Process of Education", he presents the essential components for effective learning.

| 1 | Collaborative learning |
|---|---|
| 2 | Active Learning |
| 3 | Reflective learning |
| 4 | Cultural learning |
| 5 | Reinforcement |

## 4.2. Constructing the Educational Evaluation Framework

In order to evaluate a VLE using the Bruner framework we need to establish what tools are provided within the VLE, which will support Bruner's fundamental principles. The issues raised by Bruner involve the following areas.

### 4.2.1. Collaborative learning
When different kinds of students are learning different things, bringing them together to solve problems more effectively to promote learning. In the manner of a beehive, different parts can be brought together to create a whole solution to a different problem. Students can do more in groups than they can on their own. Shulman (1998).

Bruner has spoken at length about the value of collaborative learning where students help each other to grasp the essence of the topic. This acknowledges the value of the interplay of peer assistance and is one of Bruner's fundamental elements of learning Bruner 1997, Bruner 1998[14]. 77% of all e-learning is designed for sole use (Galvin 2001)[15] Their evidence is that participants who studied together learn more than those who studied alone. (Johnson & Johnson 1990)[16]

In addition to Bruner's insight we have also consulted Gagne's Conditions of Learning and Vygotsky 's educational theory. The latter in particular supporting the importance of collaboration where the learner can "always do more than he can independently" Vygotsky, 1987, p.209[17]

In order to measure the degree of collaborative learning that the various VLEs allowed, a series of questions were constructed and submitted to student users. These included

*Q1* *Were discussion boards, chat areas available*
*Q2* *How useful would you rate the facilities for working with other students on-line*
*Q3* *Did you have access to other student's comments through the VLE?*
*Q4* *Please rate how effective the use of discussion boards, message boards, forums and chat areas etc. were on the VLE*

### 4.2.2. Active learning
Students who are actively seeking information, who engage in the learning process rather than passively receiving instruction learn more effectively. This is supported also by Lee Shulman, (1998)[18] who regards active learning as a more effective means of learning.

Vygotsky also lends his support to the importance of active learning when he says "to learn how to swim you have to, out of necessity plunge right into the water... so the only way to learn ... is by doing so" Vygotsky 1997 pp324[19]

Bruner (1997) included learner control as one of his primary educational principles that involves the learner being able to take control of his learning process. Lee & Lee (1991)[20] have shown that the outcomes of learner control over sequential tasks gave strong results, particularly during the initial learning phase.

The VLE has a number of techniques for implementing learner control which include the following:

- **Course menus**: allows learners to chose specific lessons and topics
- **External content links**: allows learners to link to relevant information elsewhere
- **Pop ups**: allows learner to obtain additional information without leaving the page
- **Standard navigation buttons**: activating forward, back and quit options
- **Guided tours**: Demonstrate the resources available in the course

The following questions were included in the questionnaire as a test of these principles.

**Q1** *Please rate the helpfulness of the menu system*
**Q2** *Please rate the degree of control that you felt you had in choosing the areas you wanted*
**Q3** *Please rate the usefulness of the navigational controls and links*
**Q4** *Please rate the appropriateness of the help pages and guided tours (if any)*

### 4.2.3. Reflective learning

The opportunity for students to reflect on the processes which they themselves engage in while learning is what Bruner calls "Going meta". This is the act of thinking about thinking. When students ask themselves "How did I think that out, How did I get there, what was the evidence, what did I do this time to solve the problem?" then they are assisting their own learning process. Any elements of a VLE, which assists the students to go "meta", would improve the learning environment.

The following questions were included to test the reflective learning environment of the Website.

**Q1** *Were you able to engage in group activities on-line*
**Q2** *How well do you feel the VLE promote your studies*
**Q3** *How do you rate the Website in its encouragement it gives for learning?*
**Q4** *How do you feel the Website fits into your learning style?*

### 4.2.4. Cultural Environment of learning

Learning and thinking are always situated in a cultural setting. The creation of a community or cultural environment that nourishes, sustains, houses and gives meaning to interactive learning. Culture concerns a system of values, rights, exchanges, obligations, opportunities and power. It concerns how humans come to know each other's minds and thus support their common processes.

The following questions test the cultural learning environment of the Website.

*Q1 How do you rate the Website in its encouragement it gives for learning?*
*Q2 How do you feel the Website fits into your learning style?*
*Q3 Do you feel that the VLE is sympathetically designed towards the tasks you need to accomplish?*

### 4.2.5. Reinforcement

This takes us to the stage after learning has occurred where the learner needs to establish and maintain his learning. This can be done by providing feedback about the correctness of the performance and by repetition of the knowledge gained. So as to ensure that learning takes hold, the learner can be asked to rehearse the knowledge gained and to show by some activity that he has achieved the goal intended. Gagne1977 pp 254[21]

Gagne writes about internal processes involved in learning involving such activities as gaining attention, informing learners of their objectives etc. (see appendix 1) but whereas these are suitable guidelines for lecturing they are not considered here to be educational principles as presented below. However one of his processes is, we believe, relevant, namely that of reinforcement, which supersedes learning process is relevant to be added to these principles.

*Q1 How well does the VLE reinforce your learning process?*
*Q2 Please rate the VLE according to a number of different ways that the information is provided to you*
*Q3 Please rate the VLE according to its ability to test your knowledge*
*Q4 Please evaluate the VLE with respect to the way it establishes your learning*

The following table summarises the educational evaluation framework for VLEs using the Bruner model as criteria against which to identify the tools and structuring provided by the VLE.

| | Tools | Structuring |
|---|---|---|
| **Collaboration** | What tools assist? Forums, Chat, email, shared work areas? | Can discussions easily be put together? |
| **Control** | Can the student tailor the system in a way which is personally beneficial | Is the structure amenable to flexible approaches? |
| **Culture** | How does the VLE engender a culture of learning? What tools are used for this? | Does the structure of the VLE foster a sense of learning identity? |
| **Reflection** | What tools promote reflection? Journals, Mind mapping? | How does the VLE promote meta-cognition? Learning about learning? |
| **Reinforcement** | What tools reinforce learning? | How does the VLE establish the learning in the mind of the student? |

### 4.3. The Educational Metric

The method used to create the Educational metric involves the scoring of the five Educational principles chosen by users during the online field survey and the conversion of those scores into a single figure index ranging from 1 to 10.

Each of the five Educational principles was evaluated by using a number of distinct questions for each principle as follows

| | Educational Principle | Number of Questions |
|---|---|---|
| 1 | Collaborative learning | 4 |
| 2 | Learner control | 4 |
| 3 | Reflective learning | 2 |
| 4 | Cultural learning | 3 |
| 5 | Reinforcement | 4 |
| | **TOTAL** | **17** |

Each question had a score rated from 1 to 5. The number of questions for each principle varies as they reflect the weighting given to them.

Stage one involved taking the average score for total of the seventeen questions. Stage two involves multiplying that score by a factor of 2 to obtain a result out of ten.

**EDI index = Average (Likert score for each EDI Question) x 2**

In the case of relative comparison of two VLEs the Likert score is again applied uniformly to each of the answers. This is possible because all the comparative answers are identical and the same formulae will follow. The score then follows the same interpretation as the HCI score.

## 5. Choosing and Setting up the Open Source VLE

A number of open source VLEs were available for use including Clara Online, Moodle and Atutor. These were each examined in detail and their strength and weaknesses considered.

| Clara Online | Strengths | Open Source, reasonable sized user base. |
|---|---|---|
| | Weaknesses | Infrequent updates and slow development. Not as fully customisable as other VLEs. |
| Moodle | Strengths | Largest open source VLE with the most active-working participants. Large user forum with dozens of posts daily. Most frequent updates and development. Rapidly becoming the most widely used open source VLE listed in 192 countries. Highly customisable through php. Fully comprehensive third party instruction manual. Multi-themed interface available. Usable with standard packages such as 'hot potatoes'. In addition Moodle was developed by its author on the basis of corresponding to a social constructivist philosophy and established on well-accepted educational principles. |
| | Weaknesses | Indeterminate |
| Atutor | Strengths | Attractive and interesting user interface |
| | Weaknesses | After a promising start Atutor has slipped behind in popularity and does not have the wide user base of other open source VLEs. |

It was decided on the basis of the information above to adopt the Moodle VLE as the most appropriate open source vehicle to use for comparison.

### 5.1. Creation of the Moodle VLE

The Moodle VLE required a web host that can support php and a MYSQL database. A number of web host companies were considered and QAwebhosting was chosen as meeting all the required criteria at a reasonable price. A domain name www.vita-uk.co.uk was also bought to enable a clear identification to be made.

The set up of the VLE was fully described in the online Moodle manual and involved the construction of a MYSQL database to store all VLE data. Installation was through the running of a number of php scripts, which installed the relevant files in the correct directories and attached connections to the MYSQL database. A configuration file needed to be set to point to the appropriate locations and index pages, which were created to provide the appropriate links. Much of the process was automated via the software provided with Moodle

The moodle.org site has a number of helpful forums and has engendered a very lively and helpful community of developers who are willing to assist anyone in the application of the open source VLE. Recourse was made to this forum on occasion, particularly during the installation when error messages were thrown up concerning the blocks site menu. By posting a message on this forum (see Appendix 2) replies were received including help from the creator of Moodle Martin Dougiamas, explaining the nature of the problem that was then easily remedied. Engaging with the development community was a straightforward and valuable process. See Appendix 2.

## 6. Overview of the Moodle and Blackboard VLEs

### 6.1. Moodle
Moodle is an open source VLE, which was constructed out of the PHD work of Martin Dougiamas. The educational philosophy upon which Moodle is based is Social Constructivism and Moodle includes tools, which are specific to constructivist learning. This places the social world of the learner as the fundamental element in learning. The VLE provides the perfect environment with its forums email communications feedback mechanisms and its ability to deal with "social entities" as learners e.g. peer groups, business teams etc. It is the interaction over time between learners, which fosters and strengthens the entire learning process through a reciprocal spiral relationship. (Dougiamas 1998)

### 6.2. Structure
Moodle offers three different formats; a weekly format that presents topics in a chronological order, a topic format which presents information thematically and a social oriented format based upon group interactions. Resources include forums, journals, quizzes, assignments, surveys, choices, chat, workshops, user profiles etc. One of the most basic elements is the forum, which is used for interactive comments between teacher and students. There are also the possibilities of adding surveys and workshops that are peer-assessed.

#### 6.2.1. Learning Activities
Moodle allows connected discussions between students with evaluation and rating of comments. There is also the use of reflective journals, glossary and encyclopedia writing, chatting, peer-evaluated assignments as learning activities. Each student has an activity report so that the lecturer can see what they have read and posted all in context. There is also a grade book. All actions are tagged with the person's photograph (optional) and name. Teachers can therefore be fully reminded of the students they are working with and can communicate with them easily through the VLE. Students are able to manage their own profile.

#### 6.2.2. Time Management and Planning
Moodle provides a calendar to include relevant dates and provide reminders of assignments and course events. Students can monitor their own learning using the activity report and provide feedback on the quality of the unit. Because Moodle is open source it is customizable and affordable and allows individual lecturers or departments to set up their own e-learning courses.

### 6.2.3. Pedagogical Approach

The constructivist philosophy is catered for with the provision of specific constructivist tools. These include the reflective journal, the peer group assessments, the rating of forum posts and a built in feedback mechanism based upon the "connected versus separate knowing" approach. Moodle contains a variety of features designed to encourage specific pedagogical practice in a simple to navigate environment.

### 6.2.4. Critique

Being an open source product it necessarily suffers from haphazard and unplanned development contingent upon the interests of its user group. However, this would avoid exposure to vendor lock-in and allows an institution to contribute to the development of their VLE in a personalised way and facilitates a move towards the adoption of a leadership role in the field if desired. Moodle does not offer any program view of courses nor does it currently afford students the capability of organising themselves or maintaining a presence outside the VLE. Some have perceived that the most critical weakness with Moodle was the limited adoption of the product on an enterprise level by any University institution. However, the situation has changed over the past couple of years and a number of FE and HE colleges have adopted Moodle as "the top ranked candidate for an institution wide deployment" (Dublin City University). In the FE sector Cornwall College, the largest college in the UK with 92000 students, has now adopted Moodle as their official VLE. Furthermore the London regional area of JISC has given Moodle a seal of approval by setting up its own Moodle site. A further issue that has been raised is that the adoption of Moodle may constrain users to its own inherent pedagogical model of social constructivism. This may be considered to be relatively unimportant, as those specific tools need not be used if they are not required.

## 6.3. Blackboard

Blackboard traces its technological roots to Cornell University. The Blackboard e-educational suite consists of three systems that make up an enterprise level e-learning platform. These are the Learning System, Portal System and Content Management System. It provides a typical model supporting a content-based transmission model that is at present the dominant mode of education.

### 6.3.1. Structure

Blackboard allows for hierarchical and sequential structuring. The course site can be broken up into group sites and each group site can have its own discussion boards, chat rooms, file exchange, and areas for project work.

### 6.3.2. Learning Activities

The generic tools included with Blackboard include discussion boards, chat rooms, assessments, grade books, announcement editors, virtual institutions to promote individual look and feel and the ability to add students. A major theme in the development of Blackboard is that of evaluation and the VLE provides evaluation by faculty, by program, by student over time and across students. Blackboard has specifically mentioned that they intend to include blogging tools (public online journal) to allow students to develop their own commentary on the course.

### 6.3.3. Time Management and Planning

Blackboard has the ability to provide a calendar, tasks and notes. It also has the ability to maintain a personal portfolio into which the user can store their various documents. However, there is no activity log, which provides a report of what a student has done during the day.

### 6.3.4. Pedagogical Approach

Blackboard does not have any specific pedagogical philosophy. "We don't seek to implement any one or multiple pedagogical models, we are agnostic in that sense"

http://www.jiscinfonet.ac.uk/InfoKits/effective-use-of-VLEs/intro-to-VLEs/i

### 6.3.5. Critique

Blackboard does not do a solid job of connecting different things in a way that is instructionally useful. A spokesman from blackboard has admitted that "My biggest area of critique of our product.... is that the data we expose of what is happening in the learning environment tends to be fairly trivia oriented..."

http://www.jiscinfonet.ac.uk/InfoKits/effective-use-of-VLEs/intro-to-VLEs/i

Blackboard is also too strongly a top down model. However recent versions are seeing an increase in the bottom up approach allowing instructors to build their own environments which complement the tools provided for centrally managed programs. Blackboard is equally candid about the fact that current tracking facilities in their product fall short of the mark in providing useful information but this is under development.

## 6.4. Summary Table

| | Moodle | Blackboard |
|---|---|---|
| **Structure** | Three formats, weekly, topical and social | Hierarchical and sequential structuring |
| **Learning Activities** | Connected discussions, Evaluation and rating of comments, Reflective journals, glossary, Encyclopaedia, writing, Chatting, peer-evaluated assignments | Discussion boards, chat rooms, assessments, grade books, announcement editors, virtual institutions |
| **Time Management and Planning** | Calendar, Activity Report | Calendar, Tasks and Notes, Portfolios |
| **Pedagogical Approach** | Constructivist philosophy | None |
| **Critique** | Haphazard and unplanned development, avoid exposure to vendor lock-in, Moodle does not offer any program view of courses nor does it currently afford students the capability of organising themselves or maintaining a presence outside the VLE | Does not connect things in an instructionally useful way. "Learning environment is trivia orientated" Tracking facilities doe not provide useful information. |

## 7. Analysis of Results

The results of the survey are purely comparative in that all the questions asked of the participants ask them to rate which VLE scored the best or worst. This meant that if Moodle scored high Blackboard automatically scored low and vice versa. This means that the results for Moodle and Blackboard are strongly correlated, and this fact needs to be borne in mind when interpreting the results below.

Some attempt is also made in this section to disentangle the results and provide individual scores for each VLE, and this is discussed later.

The Survey was embedded into Moodle, which performed a number of automated tasks as follows.

The first analysis provided the number of responses for each individual answer to the 34 survey questions. This item response analysis determined the relative strengths and weaknesses of each VLE in any of the five Educational and eight HCI areas. See Results Table 1 – BASE DATA (Appendix 4)

In the second analysis an automatic list was generated with the name of each participant together with their overall percentage grade comparing the two VLEs. This crude overall grade gave a very simple macro view of the relative perceived merits of each VLE. See Results Table 2 – BASE DATA (Appendix 4)

The third analysis performed automatically was a listing of items in the survey with summary statistics. This provided the percentage breakdown of answers for each question enabling a relative determination of the strength of each VLE down to the individual question level. See Results Table 3 – BASE DATA (Appendix 4)

The automatic facilities further provided the downloading of those statistics in excel format for further analysis.

### 7.1. Specific Educational Results

The initial result showed that the Blackboard VLE scored 5.1, which was fractionally more the Moodle VLE at 4.9 in terms of educational values. This is a difference by a factor of 2% and suggests that within the accuracy of the study Moodle and Blackboard equally meets the educational needs of the students.

Blackboard scored more highly in the areas of learner control (+2.25%) and reflective learning (+0.75%). Although not as significant as the HCI differences they nevertheless indicate that they had more control over the learning process with Blackboard than with Moodle and that Blackboard enables them to think more about the process of learning.

The only area where Moodle scored more highly was in the cultural environment of learning (+0.9%), which suggests that Moodle fosters a holistic cultural identity in a slightly more effective way than Blackboard.

The frequency analysis, see results table 5, (Appendix 4) for the Educational index indicates slight skewing in favour of Moodle as opposed to Blackboard.

### 7.2. Summary of Specific individual Educational results

Moodle fared best with
Cultural learning  +0.9%

Blackboard fared best with
Learner control + 2.25%
Reflective learning + 0.75%
Reinforcement + 0.25%
Collaborative learning + 0.25%

### 7.3. Specific HCI Results

The second result showed that the Moodle VLE scored 4.7 as against Blackboard's score of 5.3 in the area of HCI implementation. This is a difference of 6% and suggests that within the accuracy of the study Blackboard and Moodle equally meet the usability needs of the students.

Blackboard scored most highly in the areas of task migratability (+ 16%) and responsiveness (+9%), whereas Moodle most strongly featured dialogue initiative (+3%). This suggests that Blackboard is perceived as the most efficient system for providing a high rate of communication between the user and the system. In addition to this the high task migratability score suggests that users find that they have more control over the system. On the other hand the slightly elevated score for Moodle in terms of dialogue initiative suggests that users felt that the dialogue interface was more effective and helpful.

The frequency analysis see results table 4 for HCI gives a slightly more interesting picture indicating a slight skewing towards the Moodle end when the weighting factors are not taken into account. This shows that apportioning of percentages to the various answers might have an effect upon the overall result. The simple frequency analysis of the numbers of questions answered shows a slight favour in terms of Moodle.

### 7.4. Summary Specific individual HCI results

Moodle fared best with
Dialogue initiative +2.5%
Error recovery +1%
Substitutivity +.0.3%
Familiarity +.0.3%
Consistency +0.25%

Blackboard fared best with
Task migratability +16%
Responsiveness +9%
Customisability +2%^

### 7.5. Section on errors

Question 31 and Question 34 were identical namely "Which VLE was more helpful in reinforcing your learning process" This was included as a simple test of consistency and it was interesting to discover that the answers for Question 31 were equally divided between Moodle and Blackboard while the answers to question 34 were slightly in favour of Moodle. This may show the degree of credibility, which can be place on this small sample of answers.

Specifically question 31 scored 46.9% whereas question 34 scored 59.4%. This represents a difference of about 12%, which suggests that a margin of error needs to be considered. However, the interpretation and implications of this are not straightforward.

## 8. Discussion and Conclusions

The size of the sample was 8 participants. This is at the mid range of what is considered an acceptable sample size by Dumas & Reditch 1999 who recommended five to 12 testers as being enough. However the more testers there are the more representative the findings will become. It was felt that this survey would have benefited from a larger sample to enable the analysis to have a more complete representation of student numbers.

### 8.1. Educational Principles

The results showing that Blackboard had a stronger score in the area of educational principles of 51% indicates that this site more closely corresponds to Jerome Bruner's Framework for Constructivism. This may be a surprising result as Martin Dougiamas, is a student of constructivism and claims to have built Moodle around this framework. What these results show is difficult to determine because of the closeness of the small variation of 2%. It may well be considered that both VLEs are as good as each other in the area of Educational Principles. It might have been expected that Moodle would have scored more highly in this area because it was deliberately constructed along these constructivist principles. This was not the case, which indicates that the incorporation of a particular educational philosophy might not be as important as adopting sound practical usage. It may need to be questioned whether constructivist Moodle is as constructivist as its designer thinks it is.

### 8.2. HCI Principles

The results here showed that Blackboard had a stronger score for its implementation of HCI principles, which made it the more usable VLE as far as the students were concerned. This is an important result as usability plays a large part in the learning tool and the process of teaching and learning. The perfect VLE would be entirely transparent to the user who would perceive only learning objectives and have unrestricted access to them.

## 8.3. Lessons to be learnt

Both VLEs are perceived by this survey to contain weaknesses in certain areas, which could benefit from further consideration by the creators. This includes attention to consistency in use and recoverability from mistakes and feedback information via dialogue boxes and ease of understanding and providing more choice in the way to accomplish various tasks. These are issues, which are primarily HCI considerations, and the suggestion is that Moodle is slightly weaker in its implementation of HCI principles.

Moodle is perceived by this survey to contain weaknesses in the areas of both familiarity of use and expected responses. This is not an unsurprising result as all of the users in the survey had continuous and persistent use of the Blackboard VLE over a number of years whereas their introduction to the Moodle VLE was in most cases expected to be brief and new. It would therefore be expected that in the area of familiarity Blackboard would score more highly with this sample of users. As such this result cannot be taken as establishing anything significant. In order to provide a significant result it would be required to find groups of users who had no previous experience of either VLE in question.

## 8.4. Personal lessons

Difference between the relative and absolute approaches need further analysis. Evaluating one Website against another is qualitatively different from analysing a Website on its own and the relevance of the former over the latter may not be as valuable as first thought.

The process of setting up an alternative VLE was a significant challenge and provided an insight into the operational development of these educational structures, but particularly the role of the teacher in setting up, operating and monitoring a VLE from behind the scenes. A whole different realm of research and analysis could have been undertaken from the back office point of view rather than the shop front. This would have meant looking at the facilities and operations of the VLE from the Lecturers point of view and the level of involvement required to implement courses and manage student learning. It was felt that this was beyond the scope of this project because it would open up a much more substantial research horizon, and involve the time and energies of surveying lecturers, who might not be easily available.

## 9. Appendices

**Appendix 1**

**Frequency Analysis of HCI Principles by Authors.**

The most highly cited authors were analysed for their proposed principles and the results are displayed in the form of an Excel Spreadsheet which is not incorporated into this report but added to it as a separate paper copy because of its size.

**APPENDIX 2**

**General problems** -> **Block error message**
by vita hinze-hoare -

I have recently started seeing the following error message at the top of the general admin page

**Block site_main_menu: /home/vitaukc/public_html/moodle/blocks/site_main_menu/block_site_main_menu.php was not readable**

**Does anybody know how to fix this?**

---

**Support opportunity: 1.3.1 site_main_menu error message.**
by Ray Lawrence

(I'm not going to call them problems anymore).

I've upgraded to 1.3.1 and get this message on the admin page on my laptop and web host installations:

**Block site_main_menu: c:\program files\easyphp1-7\www\moodle\blocks/site_main_menu/block_site_main_menu.php was not readable**

**site_main_menu: This folder is readable on both sites.**

**block_site_main_menu.php: this file is not present**

Reply

Rate... ▼

**Re: Support opportunity: 1.3.1 site_main_menu error message.**
by Martin Dougiamas

Oo goody - support opportunity!

How did you upgrade?

blocks/site_main_menu has never been in Moodle 1.3.1 or earlier. So I suspect you've taken a detour through 1.4 development terrritory (not good).

Show parent | Reply

Rate... ▼

**Re: Support opportunity: 1.3.1 site_main_menu error message.**
by Martin Dougiamas

Arg, my apologies ... my new script that creates the packages from CVS didn't delete that empty directory in the package... you can just delete the **blocks**/**site_main_menu** directory in your installations to fix this message, and I'll re-roll the archive.

The nightlies are re-done, and I've just re-uploaded the 1.3.1 release to Sourceforge.

Thanks for the feedback.

# APPENDIX 3

| HCI | Customizability | 1 | Overall: Which site do you prefer to work with best? |
|---|---|---|---|
| Educational | Learner control | 2 | Which VLE had the better navigational controls and links? |
| Educational | Collaborative learning | 3 | Which VLE provided better access to community facilities ? |
| HCI | Dialogue Initiative | 4 | Which site had the most informative dialogue boxes? |
| Educational | Cultural learning | 5 | Which VLE was more sympathetically designed towards the tasks you needed to accomplish |
| HCI | Familiarity | 6 | Which site gave you the responses you expected? |
| Educational | Learner control | 7 | Which VLE had the most useful help pages and guided tours, if any |
| HCI | Recoverability | 8 | Which site allowed you to recover more easily from mistakes? |
| HCI | Familiarity | 9 | Which VLE do you think most students would use if given the choice? |
| Educational | Reinforcement | 10 | Which VLE was most able to test your knowledge |
| HCI | Task Migratability | 11 | Which site was the easiest to use for unfamiliar tasks? |
| Educational | Reinforcement | 12 | Which VLE had the most number of ways for providing information to you |
| HCI | Consistency | 13 | Which site seemed more consistent in use? |
| Educational | Learner control | 14 | Which VLE had the more helpful menu system? |
| Educational | Learner control | 15 | Which VLE gave you the better degree of control? |
| Educational | Collaborative learning | 16 | Which VLE had the better facilities for working with other students on-line? |
| HCI | Consistency | 17 | Which site would you least want to change to make it more usable? |
| HCI | Customizability | 18 | Which site is the most easily customizable? |
| HCI | Consistency | 19 | Which site was the easiest one to understand? |
| HCI | Substitutivity | 20 | Which site gave you most options for task completion? |
| Educational | Cultural learning | 21 | Which VLE better fitted in with your style of learning |
| Educational | Collaborative learning | 22 | Which VLE made better use of discussion boards, forums, message boards etc.? |
| HCI | Substitutivity | 23 | Which site gave you better choice in the way to do what you wanted? |
| Educational | Cultural learning | 24 | Which VLE encouraged learning the most? |
| Educational | Reflective learning | 25 | Which VLE did you think best promoted your studies |
| HCI | Responsiveness | 26 | Which site has the best look and feel? |
| Educational | Collaborative learning | 27 | Which VLE provided the better discussion boards, chat areas etc. |
| Educational | Reinforcement | 28 | Which VLE was better able to reinforce your learning process? |
| HCI | Familiarity | 29 | Which site worked according to your expectation? |
| Educational | Reinforcement | 30 | Which VLE was more helpful in reinforcing your learning process |
| HCI | Consistency | 31 | Which site was it easiest to find your way around? |
| Educational | Reflective learning | 32 | Which VLE   better allowed you to engage in group activities on-line |
| HCI | Familiarity | 33 | On which site did you feel most at home? |
| HCI | Recoverability | 34 | Which site gave you the least operating problems? |

**Questions and Responses**

| Q-1 | Which site was it easiest to find your way around? | | |
|---|---|---|---|
| | 0 | 0.0% | Mostly Moodle |
| | 3 | 37.5% | Moodle a little more than Blackboard |
| | 1 | 12.5% | Moodle and Blackboard equally |
| | 3 | 37.5% | Blackboard a little more than Moodle |
| | 1 | 12.5% | Mostly Blackboard |
| Q-2 | On which site did you feel most at home? | | |
| | 2 | 25.0% | Mostly Moodle |
| | 0 | 0.0% | Moodle a little more than Blackboard |
| | 1 | 12.5% | Moodle and Blackboard equally |
| | 2 | 25.0% | Blackboard a little more than Moodle |
| | 3 | 37.5% | Mostly Blackboard |
| Q-3 | Which site worked according to your expectation? | | |
| | 0 | 0.0% | Mostly Moodle |
| | 0 | 0.0% | Moodle a little more than Blackboard |
| | 4 | 50.0% | Moodle and Blackboard equally |
| | 2 | 25.0% | Blackboard a little more than Moodle |
| | 2 | 25.0% | Mostly Blackboard |
| Q-4 | Which site seemed more consistent in use? | | |
| | 0 | 0.0% | Mostly Moodle |
| | 3 | 37.5% | Moodle a little better than Blackboard |
| | 4 | 50.0% | Moodle and Blackboard equally |
| | 0 | 0.0% | Blackboard a little better than Moodle |
| | 1 | 12.5% | Mostly Blackboard |
| Q-5 | Which site allowed you to recover more easily from mistakes? | | |
| | 0 | 0.0% | Mostly Moodle |
| | 1 | 12.5% | Moodle a little better than Blackboard |
| | 6 | 75.0% | Moodle and Blackboard equally |
| | 1 | 12.5% | Blackboard a little better than Moodle |
| | 0 | 0.0% | Mostly Blackboard |
| Q-6 | Which site gave you the least operating problems? | | |
| | 0 | 0.0% | Mostly Moodle |
| | 1 | 12.5% | Moodle a little better than Blackboard |
| | 4 | 50.0% | Moodle and Blackboard equally |
| | 1 | 12.5% | Blackboard a little better than Moodle |
| | 1 | 12.5% | Mostly Blackboard |
| Q-7 | Which site had the most informative dialogue boxes? | | |
| | 0 | 0.0% | Mostly Moodle |
| | 3 | 37.5% | Moodle a little better than Blackboard |
| | 4 | 50.0% | Moodle and Blackboard equally |
| | 1 | 12.5% | Blackboard a little better than Moodle |
| | 0 | 0.0% | Mostly Blackboard |
| Q-8 | Which site was the easiest one to understand? | | |
| | 0 | 0.0% | Mostly Moodle |
| | 5 | 62.5% | Moodle a little more than Blackboard |
| | 3 | 37.5% | Moodle and Blackboard equally |
| | 0 | 0.0% | Blackboard a little more than Moodle |
| | 0 | 0.0% | Mostly Blackboard |
| Q-9 | Which site gave you better choice in the way to do what you wanted? | | |
| | 0 | 0.0% | Mostly Moodle |

|   |   |   |
|---|---|---|
| 3 | 37.5% | Moodle a liitle more than Blackboard |
| 5 | 62.5% | Moodle and Blackboard equally |
| 0 | 0.0% | Blackboard a little more than Moodle |
| 0 | 0.0% | Mostly Blackboard |

**Q-10** Which site gave you most options for task completion?

|   |   |   |
|---|---|---|
| 0 | 0.0% | Mostly Moodle |
| 2 | 25.0% | Moodle a little more than Blackboard |
| 4 | 50.0% | Moodle and Blackboard equally |
| 1 | 12.5% | Blackboard a little more than Moodle |
| 1 | 12.5% | Mostly Blackboard |

**Q-11** Which site gave you the responses you expected?

|   |   |   |
|---|---|---|
| 0 | 0.0% | Mostly Moodle |
| 0 | 0.0% | Moodle a little more than Blackboard |
| 5 | 62.5% | Moodle and Blackboard equally |
| 1 | 12.5% | Blackboard a little more than Moodle |
| 1 | 12.5% | Mostly Blackboard |

**Q-12** Which site was the easiest to use for unfamiliar tasks?

|   |   |   |
|---|---|---|
| 0 | 0.0% | Mostly Moodle |
| 3 | 37.5% | Moodle a little more than Blackboard |
| 3 | 37.5% | Moodle and Blackboard equally |
| 1 | 12.5% | Blackboard a little more than Moodle |
| 1 | 12.5% | Mostly Blackboard |

**Q-13** Which site is the most easily customizable?

|   |   |   |
|---|---|---|
| 0 | 0.0% | Mostly Moodle |
| 0 | 0.0% | Moodle a little more than Blackboard |
| 7 | 87.5% | Moodle and Blackboard Equally |
| 0 | 0.0% | Blackboard a little more than Moodle |
| 0 | 0.0% | Mostly Blackboard |

**Q-14** Which site has the best look and feel?

|   |   |   |
|---|---|---|
| 3 | 37.5% | Mostly Moodle |
| 1 | 12.5% | Moodle a little more than Blackboard |
| 0 | 0.0% | Moodle and Blackboard equally |
| 2 | 25.0% | Blackboard a little more than Moodle |
| 2 | 25.0% | Mostly Blackboard |

**Q-15** Which site would you want to change least to make it more usable?

|   |   |   |
|---|---|---|
| 0 | 0.0% | Mostly Moodle |
| 4 | 50.0% | Moodle a little more than Blackboard |
| 1 | 12.5% | Moodle and Blackboard equally |
| 1 | 12.5% | Blackboard a little more than Moodle |
| 2 | 25.0% | Mostly Blackboard |

**Q-16** Which site do you prefer to work with best?

|   |   |   |
|---|---|---|
| 0 | 0.0% | Mostly Moodle |
| 3 | 37.5% | Moodle a little more than Blackboard |
| 3 | 37.5% | Moodle and Blackboard equally |
| 0 | 0.0% | Blackboard a little more than Moodle |
| 2 | 25.0% | Mostly Blackboard |

**Q-17** Which VLE better allowed you to engage in group activities on-line

|   |   |   |
|---|---|---|
| 0 | 0.0% | Mostly Moodle |
| 2 | 25.0% | Moodle more than Blackboard |
| 6 | 75.0% | Moodle and Blackboard equally |
| 0 | 0.0% | Blackboard more than Moodle |
| 0 | 0.0% | Mostly Blackboard |

ResultsTable1
CONT

**Q-18** Which VLE provided the better discussion boards, chat areas etc. ?

|   |   |   |
|---|---|---|
| 0 | 0.0% | Mostly Moodle |
| 2 | 25.0% | Moodle more than Blackboard |
| 5 | 62.5% | Moodle  and Blackboard equally |
| 1 | 12.5% | Blackboard more than Moodle |
| 0 | 0.0% | Mostly Blackboard |

**Q-19**  Which VLE had the better facilities for working with?

|   |   |   |
|---|---|---|
| 0 | 0.0% | Mostly Moodle |
| 3 | 37.5% | Moodle more than Blackboard |
| 5 | 62.5% | Moodle and Blackboard equally |
| 0 | 0.0% | Blackboard more than Moodle |
| 0 | 0.0% | Mostly Blackboard |

**Q-20**  Which VLE provided better access to community facilities ?

|   |   |   |
|---|---|---|
| 0 | 0.0% | Mostly Moodle |
| 1 | 12.5% | Moodle more than Blackboard |
| 6 | 75.0% | Moodle and Blackboard equally |
| 1 | 12.5% | Blackboard more than Moodle |
| 0 | 0.0% | Mostly Blackboard |

**Q-21**  Which VLE do you think most students would use if given the choice?

|   |   |   |
|---|---|---|
| 0 | 0.0% | Mostly Moodle |
| 3 | 37.5% | Moodle more than Blackboard |
| 1 | 12.5% | Moodle and Blackboard equally |
| 4 | 50.0% | Blackboard more than Moodle |
| 0 | 0.0% | Mostly Blackboard |

**Q-22**  Which VLE made better use of discussion boards, forums, message boards etc.?

|   |   |   |
|---|---|---|
| 0 | 0.0% | Mostly Moodle |
| 2 | 25.0% | Moodle more than Blackboard |
| 5 | 62.5% | Moodle and Blackboard equally |
| 0 | 0.0% | Blackboard more than Moodle |
| 1 | 12.5% | Mostly Blackboard |

**Q-23**  Which VLE had the more helpful menu system?

|   |   |   |
|---|---|---|
| 0 | 0.0% | Mostly Moodle |
| 5 | 62.5% | Moodle more than Blackboard |
| 1 | 12.5% | Moodle and Blackboard equally |
| 1 | 12.5% | Blackboard more than Moodle |
| 1 | 12.5% | Mostly Blackboard |

**Q-24**  Which VLE gave you the better degree of control?

|   |   |   |
|---|---|---|
| 0 | 0.0% | Mostly Moodle |
| 1 | 12.5% | Moodle more than Blackboard |
| 7 | 87.5% | Moodle and Blackboard equally |
| 0 | 0.0% | Blackboard more than Moodle |
| 0 | 0.0% | Mostly Blackboard |

**Q-25**  Which VLE had the better navigational controls and links?

|   |   |   |
|---|---|---|
| 0 | 0.0% | Mostly Moodle |
| 2 | 25.0% | Moodle more than Blackboard |
| 3 | 37.5% | Moodle and Blackboard equally |
| 2 | 25.0% | Blackboard more than Moodle |
| 1 | 12.5% | Mostly Blackboard |

**Q-26**  Which VLE had the most useful help pages and guided tours, if any ?

|   |   |   |
|---|---|---|
| 0 | 0.0% | Mostly Moodle |
| 0 | 0.0% | Moodle more than Blackboard |
| 6 | 75.0% | Moodle and Blackboard equally |
| 1 | 12.5% | Blackboard more than Moodle |
| 0 | 0.0% | Mostly Blackboard |

ResultsTable1
CONT

**Q-27**  Which VLE encouraged learning the most?

|   |   |   |
|---|---|---|
| 0 | 0.0% | Mostly Moodle |
| 3 | 37.5% | Moodle more than Blackboard |
| 4 | 50.0% | Moodle and Blackboard equally |
| 1 | 12.5% | Blackboard more than Moodle |
| 0 | 0.0% | Mostly Blackboard |

**Q-28**  Which VLE better fitted in with your style of learning ?

|   |   |   |
|---|---|---|
| 0 | 0.0% | Mostly Moodle |
| 4 | 50.0% | Moodle more than Blackboard |
| 3 | 37.5% | Moodle and Blackboard equally |
| 1 | 12.5% | Blackboard more than Moodle |
| 0 | 0.0% | Mostly Blackboard |

**Q-29**    Which VLE was more sympathetically designed towards the tasks you needed to accomplish?

|   |   |   |
|---|---|---|
| 0 | 0.0% | Mostly Moodle |
| 1 | 12.5% | Moodle more than Blackboard |
| 6 | 75.0% | Moodle and Blackboard equally |
| 1 | 12.5% | Blackboard more than Moodle |
| 0 | 0.0% | Mostly Blackboard |

**Q-30**    Which VLE did you think best promoted your studies?

|   |   |   |
|---|---|---|
| 0 | 0.0% | Mostly Moodle |
| 1 | 12.5% | Moodle more than Blackboard |
| 4 | 50.0% | Moodle and Blackboard equally |
| 2 | 25.0% | Blackboard more than Moodle |
| 0 | 0.0% | Mostly Blackboard |

**Q-31**    Which VLE was more helpful in reinforcing your learning process ?

|   |   |   |
|---|---|---|
| 0 | 0.0% | Mostly Moodle |
| 1 | 12.5% | Moodle more than Blackboard |
| 5 | 62.5% | Moodle and Blackboard equally |
| 2 | 25.0% | Blackboard more than Moodle |
| 0 | 0.0% | Mostly Blackboard |

**Q-32**    Which VLE had the most number of ways for providing information to you ?

|   |   |   |
|---|---|---|
| 0 | 0.0% | Mostly Moodle |
| 4 | 50.0% | Moodle more than Blackboard |
| 2 | 25.0% | Moodle and Blackboard equally |
| 2 | 25.0% | Blackboard more than Moodle |
| 0 | 0.0% | Mostly Blackboard |

**Q-33**    Which VLE was most able to test your knowledge?

|   |   |   |
|---|---|---|
| 1 | 12.5% | Mostly Moodle |
| 0 | 0.0% | Moodle more than Blackboard |
| 6 | 75.0% | Moodle and Blackboard equally |
| 1 | 12.5% | Blackboard more than Moodle |
| 0 | 0.0% | Mostly Blackboard |

**Q-34**    Which VLE was better able to reinforce your learning process?

|   |   |   |
|---|---|---|
| 0 | 0.0% | Mostly Moodle |
| 3 | 37.5% | Moodle more than Blackboard |
| 5 | 62.5% | Moodle and Blackboard equally |
| 0 | 0.0% | Blackboard more than Moodle |
| 0 | 0.0% | Mostly Blackboard |

| Name | Grade | Q-1 | Q-2 | Q-3 | Q-4 | Q-5 | Q-6 | Q-7 | Q-8 | Q-9 | Q-10 | Q-11 | Q-12 | Q-13 | Q-14 | Q-15 | Q-16 | Q-17 | Q-18 | Q-19 | Q-20 | Q-21 | Q-22 | Q-23 | Q-24 | Q-25 | Q-26 | Q-27 | Q-28 | Q-29 | Q-30 | Q-31 | Q-32 | Q-33 | Q-34 |
|---|---|---|---|---|---|---|---|---|---|---|---|---|---|---|---|---|---|---|---|---|---|---|---|---|---|---|---|---|---|---|---|---|---|---|---|
| dave . | 66 | 2 | 1 | 3 | 3 | 3 | 2 | 2 | 3 | 3 | 3 | -- | 2 | 3 | 1 | 3 | 2 | 3 | 3 | 3 | 3 | 2 | 3 | 2 | 3 | 3 | 3 | 2 | 2 | 3 | 4 | 3 | 2 | 3 | 3 |
| a a | 46 | 5 | 5 | 5 | 5 | 3 | 5 | 3 | 3 | 3 | 4 | 5 | 5 | 3 | 5 | 5 | 5 | 3 | 3 | 3 | 4 | 3 | 5 | 3 | 5 | 3 | 5 | 3 | 3 | 3 | 3 | 3 | 4 | 3 | 3 |
| Fernando Andreu | 59 | 4 | 4 | 4 | 3 | 4 | 4 | 3 | 2 | 2 | 2 | 3 | 3 | 3 | 4 | 2 | 3 | 2 | 2 | 2 | 2 | 4 | 2 | 3 | 2 | 4 | 3 | 3 | 2 | 4 | 4 | 4 | 3 | 1 | 3 |
| Gary Barnett | 67 | 2 | 1 | 3 | 2 | 3 | 3 | 3 | 2 | 3 | 3 | 3 | 2 | 3 | 1 | 2 | 2 | 3 | 3 | 2 | 3 | 2 | 3 | 4 | 3 | 3 | 3 | 2 | 2 | 3 | 3 | 3 | 3 | 3 | 2 |
| Jean Leach | 71 | 3 | 3 | 3 | 2 | 3 | 3 | 2 | 2 | 2 | 2 | 3 | 2 | 3 | 1 | 2 | 2 | 3 | 2 | 3 | 2 | 3 | 2 | 3 | 2 | 3 | 2 | 3 | 2 | 2 | 3 | 2 | 2 | 3 | 2 |
| james llewhellin | 57 | 4 | 5 | 5 | 3 | 3 | -- | 3 | 3 | 3 | 3 | 4 | 3 | 3 | 5 | 5 | 5 | 2 | 3 | 2 | 3 | 4 | 2 | 2 | 3 | 3 | 3 | 3 | 3 | 2 | 3 | 3 | 2 | 3 | 2 |
| reena pau | 52 | 2 | 4 | 3 | 2 | 2 | 2 | 4 | 2 | 3 | 5 | 3 | 4 | 3 | 4 | 4 | 3 | 3 | 4 | 3 | 4 | 4 | 5 | 2 | 3 | 4 | 4 | 4 | 4 | 3 | -- | 4 | 4 | 4 | 3 |
| amw White | 57 | 4 | 5 | 4 | 3 | 3 | 3 | 2 | 3 | 2 | 3 | 3 | 3 | -- | 2 | 2 | 3 | 3 | 3 | 3 | 3 | 3 | 2 | 3 | 2 | -- | 3 | 3 | 3 | 3 | 3 | 3 | 2 | 3 | 3 |

| VLE Survey | Perc in fav | Average | 1 | 2 | 3 | 4 | 5 | 6 | 7 | 8 | 9 | 10 | 11 | 12 | 13 | 14 | 15 | 16 | 17 |
|---|---|---|---|---|---|---|---|---|---|---|---|---|---|---|---|---|---|---|---|---|
| dave . | 8.1% | 58.09% | 0.75 | 1 | 0.5 | 0.5 | 0.5 | 0.5 | 0.75 | 0.75 | 0.5 | 0.5 | 0 | 0.75 | 0.5 | 1 | 0.5 | 0.75 | 0.5 |
| a a | -16.9% | 33.09% | 0 | 0 | 0 | 0 | 0.5 | 0 | 0.5 | 0.5 | 0.5 | 0.25 | 0 | 0 | 0.5 | 0 | 1 | 0 | 0.5 |
| Fernando Andreu | -0.7% | 49.26% | 0.25 | 0.25 | 0.25 | 0.5 | 0 | 0.25 | 0.5 | 0.75 | 0.75 | 0.75 | 0.5 | 0.5 | 0.5 | 0.25 | 0.25 | 0.5 | 0.75 |
| Gary Barnett | 8.8% | 58.82% | 0.75 | 1 | 0.5 | 0.75 | 0.5 | 0.5 | 0.5 | 0.75 | 0.5 | 0.5 | 0.5 | 0.75 | 0.5 | 1 | 0.25 | 0.75 | 0.5 |
| Jean Leach | 13.2% | 63.24% | 0.5 | 0.5 | 0.5 | 0.75 | 0.5 | 0.5 | 0.75 | 0.75 | 0.75 | 0.75 | 0.5 | 0.75 | 0.5 | 1 | 0.25 | 0.75 | 0.5 |
| james llewhellin | -2.9% | 47.06% | 0.25 | 0 | 0 | 0.5 | 0.5 | 0 | 0.5 | 0.5 | 0.5 | 0.5 | 0.25 | 0.5 | 0.5 | 0 | 1 | 0 | 0.75 |
| reena pau | -9.6% | 40.44% | 0.75 | 0.25 | 0.5 | 0.75 | 0.75 | 0.75 | 0.25 | 0.75 | 0.5 | 0 | 0.5 | 0.25 | 0.5 | 0.25 | 0.75 | 0.5 | 0.5 |
| amw White | -2.2% | 47.79% | 0.25 | 0 | 0.25 | 0.5 | 0.5 | 0.5 | 0.75 | 0.5 | 0.75 | 0.5 | 0.5 | 0.5 | 0 | 0.75 | 0.25 | 0.5 | 0.5 |
| | | | | | | | | | | | | | | | | | | | |
| % | | | 43.8% | 37.5% | 31.3% | 53.1% | 46.9% | 37.5% | 56.3% | 65.6% | 59.4% | 46.9% | 34.4% | 50.0% | 43.8% | 53.1% | 53.1% | 46.9% | 56.3% |
| Percent in favour | | | -6.2% | -12.5% | -18.7% | 3.1% | -3.1% | -12.5% | 6.3% | 15.6% | 9.4% | -3.1% | -15.6% | 0.0% | -6.2% | 3.1% | 3.1% | -3.1% | 6.3% |

| 18 | 19 | 20 | 21 | 22 | 23 | 24 | 25 | 26 | 27 | 28 | 29 | 30 | 31 | 32 | 33 | 34 |
|---|---|---|---|---|---|---|---|---|---|---|---|---|---|---|---|---|
| 0.5 | 0.5 | 0.5 | 0.75 | 0.5 | 0.75 | 0.5 | 0.5 | 0.5 | 0.75 | 0.75 | 0.5 | 0.25 | 0.5 | 0.75 | 0.5 | 0.5 |
| 0.5 | 0.5 | 0.5 | 0.25 | 0.5 | 0 | 0.5 | 0 | 0.5 | 0.5 | 0.5 | 0.5 | 0.5 | 0.5 | 0.25 | 0.5 | 0.5 |
| 0.75 | 0.75 | 0.75 | 0.25 | 0.75 | 0.5 | 0.75 | 0.25 | 0.5 | 0.5 | 0.75 | 0.25 | 0.25 | 0.25 | 0.5 | 1 | 0.5 |
| 0.5 | 0.75 | 0.5 | 0.75 | 0.5 | 0.25 | 0.5 | 0.5 | 0.5 | 0.75 | 0.75 | 0.5 | 0.5 | 0.5 | 0.5 | 0.5 | 0.75 |
| 0.75 | 0.5 | 0.5 | 0.75 | 0.5 | 0.75 | 0.5 | 0.75 | 0.5 | 0.75 | 0.75 | 0.5 | 0.75 | 0.75 | 0.75 | 0.5 | 0.75 |
| 0.5 | 0.75 | 0.5 | 0.25 | 0.75 | 0.75 | 0.5 | 0.5 | 0.5 | 0.5 | 0.5 | 0.5 | 0.75 | 0.5 | 0.75 | 0.5 | 0.75 |
| 0.25 | 0.5 | 0.25 | 0.25 | 0 | 0.75 | 0.5 | 0.25 | 0.25 | 0.25 | 0.25 | 0.5 | 0 | 0.25 | 0.25 | 0.25 | 0.5 |
| 0.5 | 0.5 | 0.5 | 0.5 | 0.5 | 0.75 | 0.5 | 0.75 | 0 | 0.5 | 0.5 | 0.5 | 0.5 | 0.5 | 0.75 | 0.5 | 0.5 |
|  |  |  |  |  |  |  |  |  |  |  |  |  |  |  |  |  |
| 53.1% | 59.4% | 50.0% | 46.9% | 50.0% | 56.3% | 53.1% | 43.8% | 40.6% | 56.3% | 59.4% | 50.0% | 40.6% | 46.9% | 56.3% | 53.1% | 59.4% |
| 3.1% | 9.4% | 0.0% | -3.1% | 0.0% | 6.3% | 3.1% | -6.2% | -9.4% | 6.3% | 9.4% | 0.0% | -9.4% | -3.1% | 6.3% | 3.1% | 9.4% |

## Appendix 4

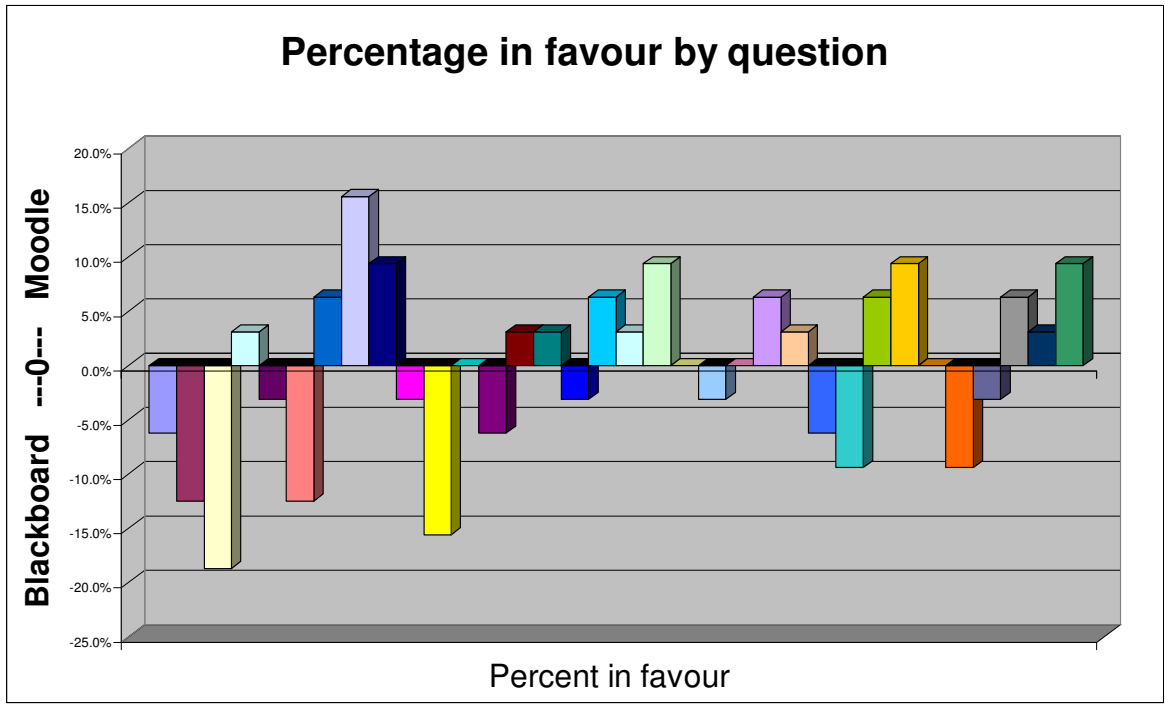

| Question Number | HCI | HCI Type | Ed Type | Ave Score | Rel ave score |
|---|---|---|---|---|---|
| 1 | HCI | Customizability | | 43.80% | -6.20% |
| 2 | Educational | | Learner Control | 37.50% | -12.50% |
| 3 | Educational | | Collaborative Learning | 31.30% | -18.70% |
| 4 | HCI | Dialogue initiative | | 53.10% | 3.10% |
| 5 | Educational | | Cultural Learning | 46.90% | -3.10% |
| 6 | HCI | Familiarity | | 37.50% | -12.50% |
| 7 | Educational | | Learner Control | 56.30% | 6.30% |
| 8 | HCI | Error Recovery | | 65.60% | 15.60% |
| 9 | HCI | Familiarity | | 59.40% | 9.40% |
| 10 | Educational | | Reinforcement | 46.90% | -3.10% |
| 11 | HCI | Task Migratability | | 34.40% | -15.60% |
| 12 | Educational | | Reinforcement | 50.00% | 0.00% |
| 13 | HCI | Consistency | | 43.80% | -6.20% |
| 14 | Educational | | Learner Control | 53.10% | 3.10% |
| 15 | Educational | | Learner Control | 53.10% | 3.10% |
| 16 | Educational | | Collaborative Learning | 46.90% | -3.10% |
| 17 | HCI | Consistency | | 56.30% | 6.30% |
| 18 | HCI | Customizability | | 53.10% | 3.10% |
| 19 | HCI | Consistency | | 59.40% | 9.40% |
| 20 | HCI | Substitutivity | | 50.00% | 0.00% |
| 21 | Educational | | Cultural Learning | 46.90% | -3.10% |
| 22 | Educational | | Collaborative Learning | 50.00% | 0.00% |
| 23 | HCI | Substitutivity | | 56.30% | 6.30% |
| 24 | Educational | | Cultural Learning | 53.10% | 3.10% |
| 25 | Educational | | ReflectiveLearning | 43.80% | -6.20% |
| 26 | HCI | Responsiveness | | 40.60% | -9.40% |
| 27 | Educational | | Collaborative Learning | 56.30% | 6.30% |
| 28 | Educational | | Reinforcement | 59.40% | 9.40% |
| 29 | HCI | Familiarity | | 50.00% | 0.00% |
| 30 | Educational | | Reinforcement | 40.60% | -9.40% |
| 31 | HCI | Consistency | | 46.90% | -3.10% |
| 32 | Educational | | ReflectiveLearning | 56.30% | 6.30% |
| 33 | HCI | Familiarity | | 53.10% | 3.10% |
| 34 | HCI | Error Recovery | | 59.40% | 9.40% |

| | | | | | | |
|---|---|---|---|---|---|---|
| Familiarity | 50.68% | 0.68% | | Collaborative Learning | 49.76% | -0.24% |
| Consistency | 50.51% | 0.51% | | Learner Control | 47.78% | -2.22% |
| Customizability | 48.28% | -1.72% | | Reflective Learning | 49.24% | -0.76% |
| Responsiveness | 40.60% | -9.40% | | Cultural Learning | 50.95% | 0.95% |
| Dialogue In | 53.10% | 3.10% | | Reinforcement | 49.71% | -0.29% |
| Task Migrate | 34.40% | -15.60% | | **TOTAL ED REL INDEX** | **49.49%** | **-0.51%** |
| Error Recovery | 51.29% | 1.29% | | | | |
| Substitutivity | 50.80% | 0.80% | | | | |
| **TOTAL HCI REL INDEX** | **47.46%** | **-2.54%** | | | | |

| Moodle | 4.7 | Blackboard | 5.3 | Moodle | 4.9 | Blackboard | 5.1 |
|---|---|---|---|---|---|---|---|

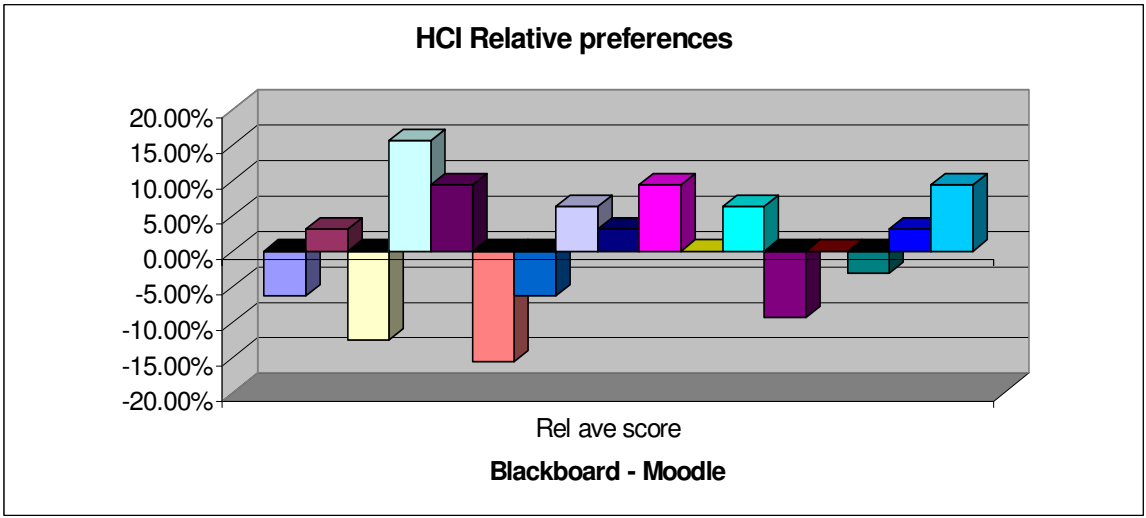

| HCI | |
|---|---|
| Question Number | Relative score |
| 1 | -6.25% |
| 4 | 3.13% |
| 6 | -12.50% |
| 8 | 15.63% |
| 9 | 9.38% |
| 11 | -15.63% |
| 13 | -6.25% |
| 17 | 6.25% |
| 18 | 3.13% |
| 19 | 9.38% |
| 20 | 0.00% |
| 23 | 6.25% |
| 26 | -9.38% |
| 29 | 0.00% |
| 31 | -3.13% |
| 33 | 3.13% |
| 34 | 9.38% |

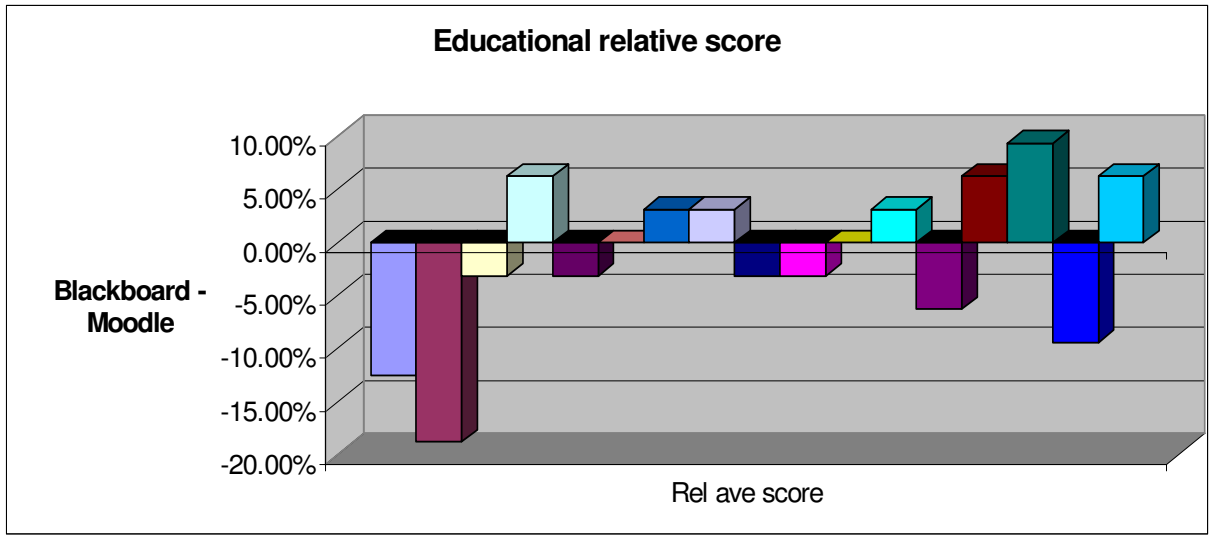

| Educational | |
|---|---|
| Question Number | Relative score |
| 2 | -12.50% |
| 3 | -18.75% |
| 5 | -3.13% |
| 7 | 6.25% |
| 10 | -3.13% |
| 12 | 0.00% |
| 14 | 3.13% |
| 15 | 3.13% |
| 16 | -3.13% |
| 21 | -3.13% |
| 22 | 0.00% |
| 24 | 3.13% |
| 25 | -6.25% |
| 27 | 6.25% |
| 28 | 9.38% |
| 30 | -9.38% |
| 32 | 6.25% |

none

**APPENDIX 4**      **Relative HCI Principles**

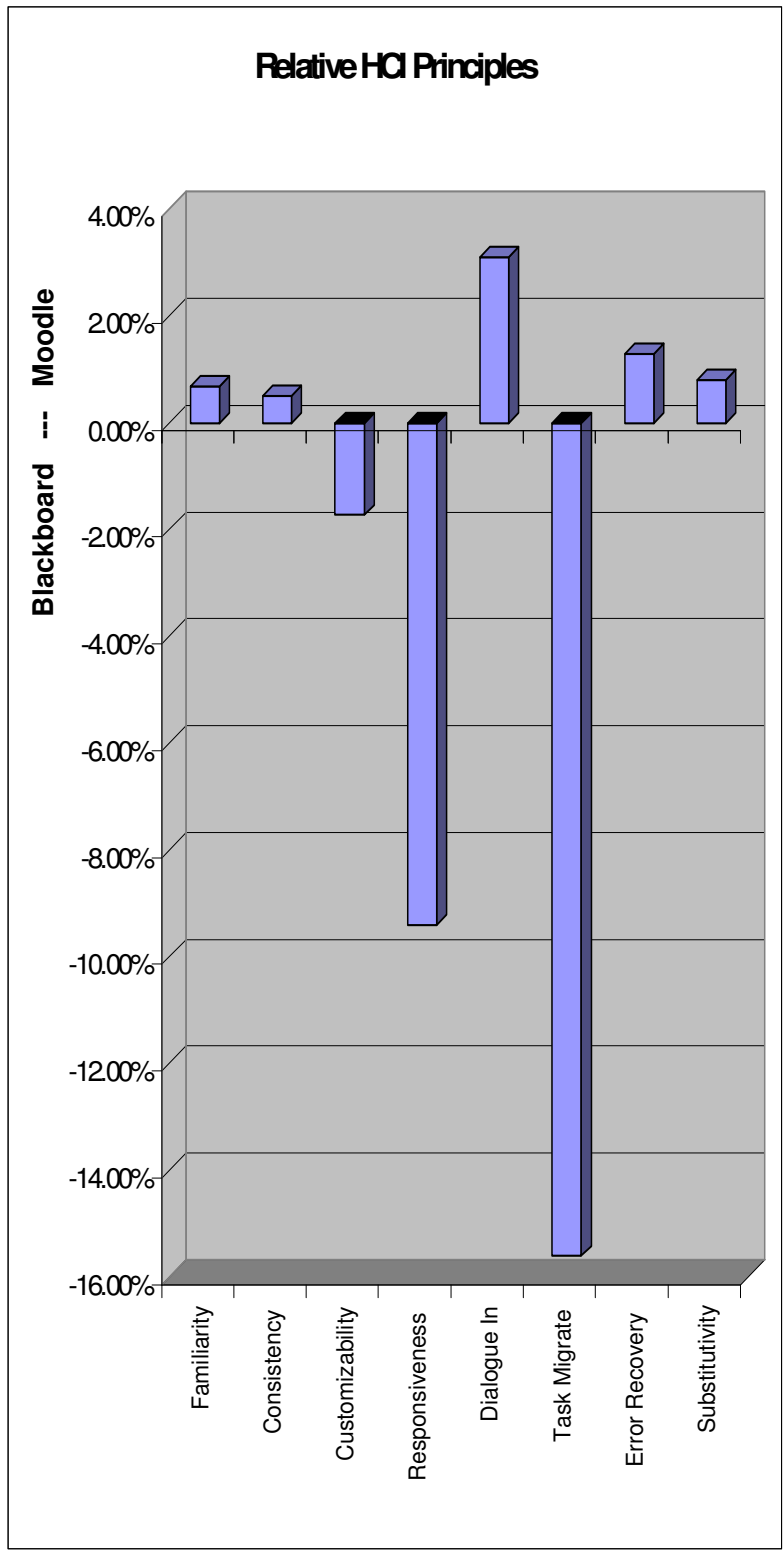

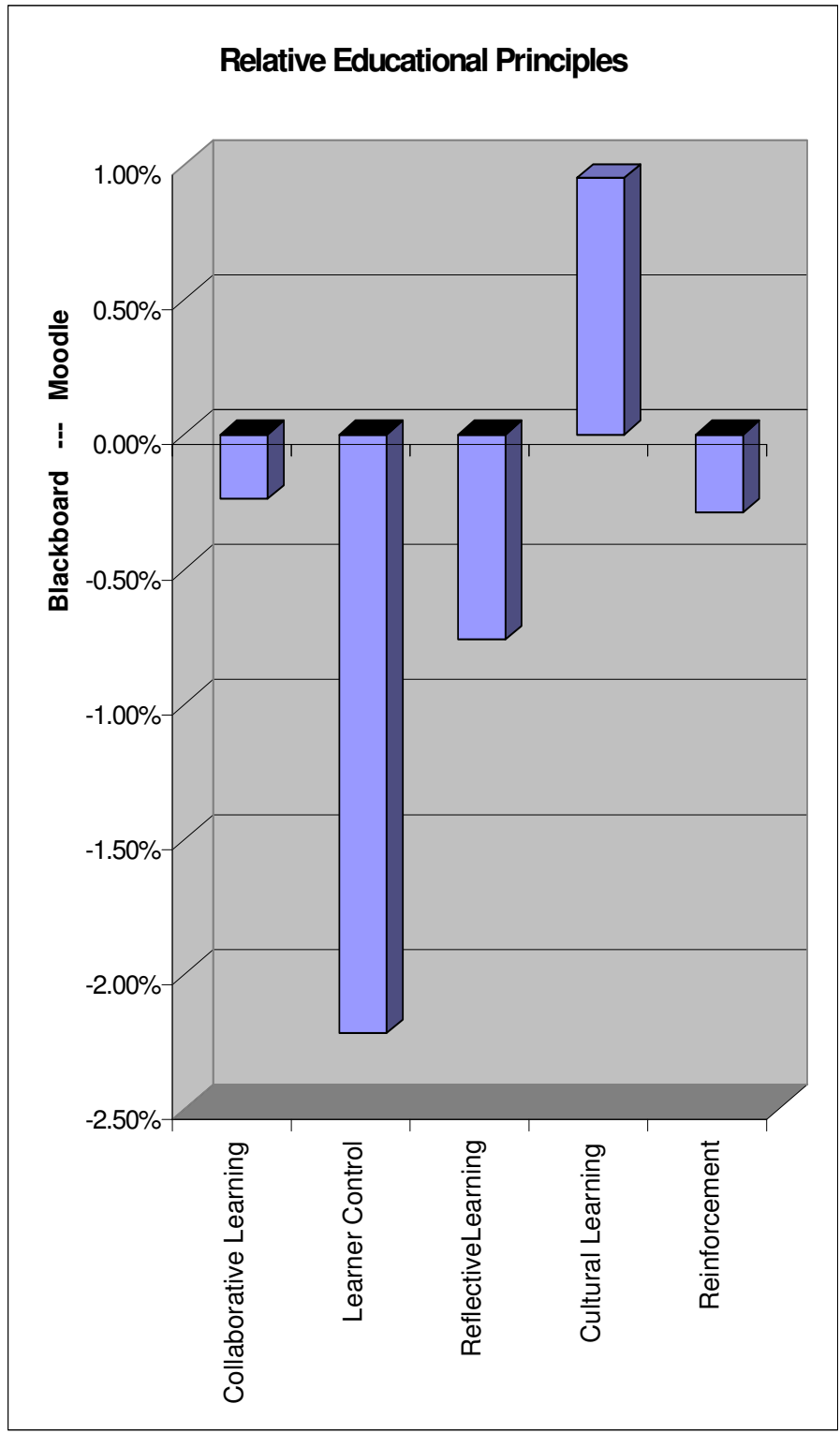

**Item Response**       **HCI**
**Analysis**

| Question | Q-1 | Q-4 | Q-6 | Q-8 | Q-9 | Q-11 | Q-13 | Q-17 | Q-18 | Q-19 | Q-20 | Q-23 | Q-26 | Q-29 | Q-31 | Q-33 | Q-34 | |
|---|---|---|---|---|---|---|---|---|---|---|---|---|---|---|---|---|---|---|
| | H | H | H | H | H | H | H | H | H | H | H | H | H | H | H | H | H | |
| **M/C #1** | -- | -- | -- | -- | -- | -- | -- | -- | -- | -- | -- | -- | -- | -- | -- | 1 | -- | **1** |
| **M/C #2** | 3 | 3 | 1 | 5 | 3 | -- | -- | 2 | 2 | 3 | 1 | 5 | -- | 1 | 1 | -- | 3 | **33** |
| **M/C #3** | 1 | 4 | 4 | 3 | 5 | 5 | 7 | 6 | 5 | 5 | 6 | 1 | 6 | 6 | 5 | 6 | 5 | **80** |
| **M/C #4** | 3 | -- | 1 | -- | -- | 1 | -- | -- | 1 | -- | 1 | 1 | 1 | 1 | 2 | 1 | -- | **13** |
| **M/C #5** | 1 | 1 | 1 | -- | -- | 1 | -- | -- | -- | -- | -- | 1 | -- | -- | -- | -- | -- | **5** |

**HCI Frequency Analysis**

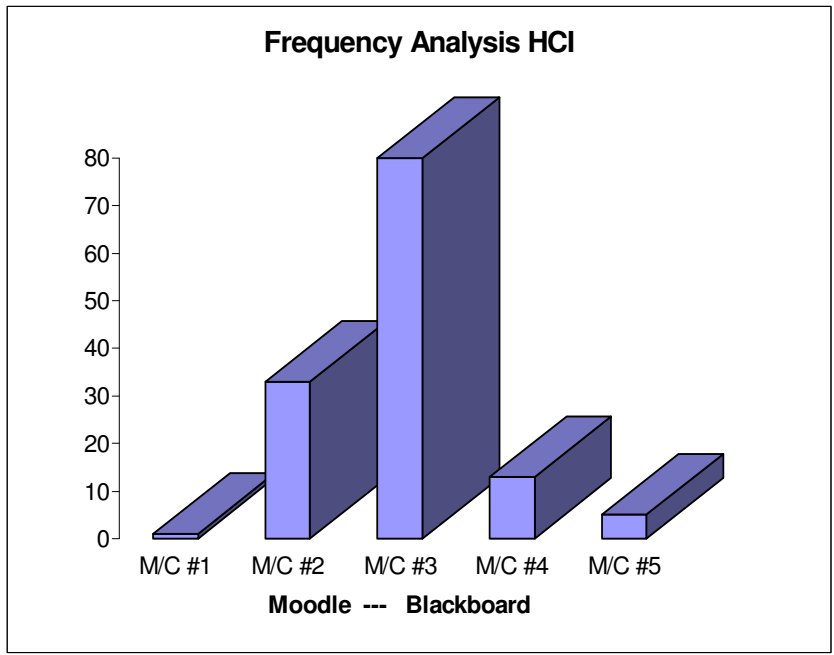

**Item Response Analysis           ED**

| Question | Q-2 | Q-3 | Q-5 | Q-7 | Q-10 | Q-12 | Q-14 | Q-15 | Q-16 | Q-21 | Q-22 | Q-24 | Q-25 | Q-27 | Q-28 | Q-30 | Q-32 | |
|---|---|---|---|---|---|---|---|---|---|---|---|---|---|---|---|---|---|---|
| **M/C #1** | 2 | -- | -- | -- | -- | -- | 3 | -- | -- | -- | -- | -- | -- | -- | -- | -- | -- | **5** |
| **M/C #2** | -- | -- | 1 | 3 | 2 | 3 | 1 | 4 | 3 | 3 | 2 | 1 | 2 | 3 | 4 | 1 | 4 | **37** |
| **M/C #3** | 1 | 4 | 6 | 4 | 4 | 3 | -- | 1 | 3 | 1 | 5 | 7 | 3 | 4 | 3 | 4 | 2 | **55** |
| **M/C #4** | 2 | 2 | 1 | 1 | 1 | 1 | 2 | 1 | -- | 4 | -- | -- | 2 | 1 | 1 | 2 | 2 | **23** |
| **M/C #5** | 3 | 2 | -- | -- | 1 | 1 | 2 | 2 | 2 | -- | 1 | -- | 1 | -- | -- | -- | -- | **15** |

**Educational Index**

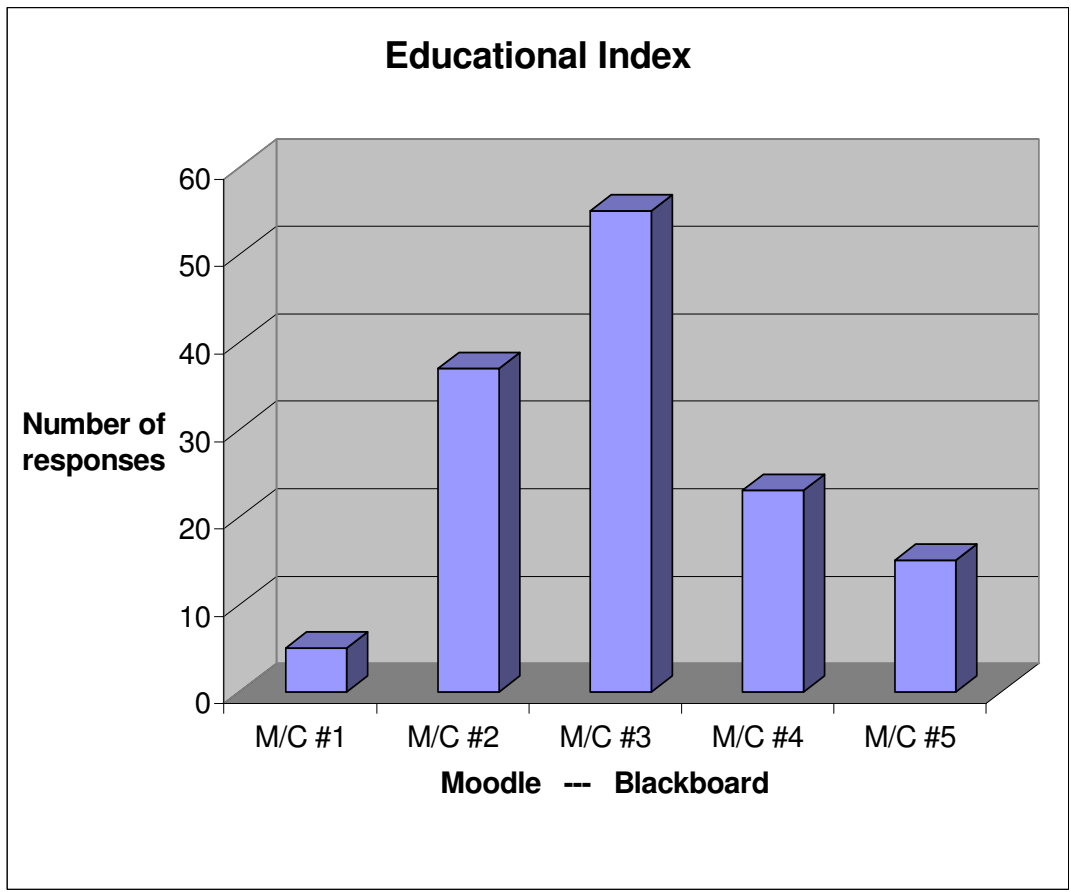

## 10.References & Bibliography

## Endnotes:

[1] Nielsen, J., (2000) *Designing Web Usability*, New Riders Publishing, P13

[2] Whittaker, S., et al  (2000) *A Reference Task Agenda for HCI*, Human-Computer Interaction in the New Millennium, John M. Carroll, Addison Wesley, 2002

[3] http://www.hi.is/~joner/eaps/dearing2.htm, http://www.leeds.ac.uk/educol/ncihe/, http://d3e.open.ac.uk/Dearing/